# Conceptual Modeling of Aggregation: an Exploration

Sabah Al-Fedaghi*

*Computer Engineering Department*
*Kuwait University*
*Kuwait*

salfedaghi@yahoo.com, sabah.alfedaghi@ku.edu.kw

*Abstract* - This paper is about conceptual modeling of aggregates in software engineering. An aggregate is a cluster of domain objects that can be treated as a single unit. In UML, an aggregation is a type of association in which objects are configured together to form a more complex object. It has been described as one of the biggest *betes noires* in modeling. In spite of its widespread use, aggregation seems a troublesome concept. It is sometimes treated as *part-of*; however, the *part-of* relation is only one of many possible aggregation mechanisms and is itself problematical, partly because of its diverse semantics. The purpose of this paper is to develop a semantic assembly model that is useful to represent relationships in the domain and achieve various levels of interoperability among software. The paper contributes to ontological conceptual clarity about aggregation, based on a model called a thinging machine (TM). The TM model uses so-called thimacs (things/machines) as building blocks for describing the domain. Thus, the notion of aggregation is extended to behavioral aggregation, in which individual entities collectively behave as a unit. The results point to a promising contribution to the understanding of the notion of aggregation compared to the ontological positions that are based on substance or relation.

## I. Introduction

A model is an interpretation of reality that abstracts the aspects relevant to solving the problem at hand and ignores extraneous detail. A *conceptual* model is an artificial creation that captures how we choose to think about the domain as we select entities and *relations* among them [1]. In software development, the *domain* is the realm of activity that is of interest to the users of the software system being designed. The domain may be a part of the real world or may be more abstract. It is "what the software is about" [1]. *Relations* among domain entities in modeling refer to associations among the objects and classes. In set theory, relations are associations of elements of sets.

Creating a conceptual model is a difficult task, and "any software project tackling complex business problems without a sharp focus on the domain model is at risk" [1]. According to Catossi et al. [2], "UML on its own is not able to guarantee a model that is free of conceptual mistakes. On trying to improve this situation techniques based on ontological analysis have been created to help modelers validate their UML class diagrams more easily."

---

*Retired June 2021, seconded fall semester 2021/2022

Abstraction has been suggested as the most common and powerful tool to help create conceptual models. Abstraction is usually considered one of the most important tools in handling the complexity of building a model. This complexity is related to the number of entities and relationships among objects. Many levels of abstraction can be identified in the modeling process, e.g., classification and aggregation. This paper is about the notion of aggregation in conceptual modeling in software engineering and database systems. The notion of aggregation is useful when one thinks about conceptual entities that are somehow constituted from smaller, self-contained entities [3].

### A. About Aggregation

"Aggregation" is a general term for any whole-part relationship among entities. The part-whole relation is, if not the, an essential relation in ontology—although its full potential is yet to be discovered [4]. In philosophy and related fields, the study of parts that compose a whole falls under the discipline of *mereology*. As a sample issue in this context, Leśniewski [5] claimed that any collection of things, e.g., the members of a set in symbolic logic, can be considered the parts of a whole and therefore can compose an "entity." Critics of this idea say that such arbitrary collections are just "scattered entities." A mind-independent connection between entities is needed for them to be integral parts [5]. Aggregation is variously defined as collection into a sum, a collection of particulars and several things grouped together or considered as a whole. It refers to a structural technique for building a new object from existing objects that support required interfaces [3]. Fowler [6] defines an aggregate as "a cluster of domain objects that can be treated as a single unit. An example may be an order and its line-items, these will be separate objects, but it's useful to treat the order (together with its line items) as a single aggregate."

Two subsets of association are aggregation and composition; however, the composition is a subset of the aggregation relationship, i.e., association ⊃ aggregation ⊃ composition. UML includes two types of aggregates: weak aggregation, e.g., players may participate in many teams at the same time; and strong aggregation (composition), e.g., chapters belong to a specific book [7]. Composition is a stronger variety of aggregation, where the part object may belong to only one whole and the parts are usually expected to live and die with the whole. Usually, any deletion of the whole is considered to cascade to the parts [8]. With composite aggregations, the part is existentially dependent on exactly one whole whereas for *shared* aggregation (the parts can be shared by many containers), there is no constraint (shared feature that must hold in all cases) on the multiplicity [5].

Central restrictions in UML-class diagrams are the aggregation/composition constraints that specify whole-part relationships between an object (the assembly) and its parts (the components) [9]. According to Bock [10], the UML 2 recognizes that an object's "parts" are best identified by navigating from the containing object along association ends or attributes to the contained objects. Links between parts are modeled as connections between the containing object's association ends.

*B. Difficulty Regarding Aggregation*

In spite of its widespread use, aggregation seems a troublesome concept. According to Veres [3], aggregation is sometimes treated as *part-of* (e.g., UML aggregation); however, the *part-of* relation is only one of many possible aggregation mechanisms and is itself problematical, partly because of its diverse semantics. In UML, the semantics of aggregation has not been made clear [3]. According to Fowler and Scott [11], "the difficult thing is considering what the difference is between aggregation and association." Fowler [8] described aggregation as "one of my biggest *betes noires* in modeling."

*C. This Paper*

The purpose of this paper is to explore the possibility of expounding the notion of aggregation in the context of conceptual modeling analysis in software and systems engineering. A possible outcome of this undertaking is developing a "semantic assembly model" [12], which is useful to represent relationships in the domain and achieve various levels of interoperability among future software [12-13]. Specifically, the paper contributes to the ontological clarity about *aggregation*, based on a model called a thinging machine (TM) [14-15]. The TM model uses so-called thimacs (*thi*ngs/*mac*hines) as building blocks for describing the domain. A thimac is a thing and simultaneously a machine. It represents a distinct entity in a domain, whether a whole or part. The machine takes five generic actions: create, process, release, transfer and receive. A thimac's behavior is rooted in these five actions.

A thimac has a minimum level of organization with respect to the whole-part relationship. The organization in this context refers a set of *constraints* on the actions and flows of things in the thimacs and their subthimacs. Thimacs have a hierarchical relationship with each other. Every thimac is something whole, and all thimacs are essentially parts of some thimac, which in turn is a subthimac of a bigger thimac.

According to Wheatley [16], at any given level of reality, a *unit's thingness* has two aspects: wholeness and parts, such that the wholeness "envelopes" its parts such that "parts are made to be pierced together into wholes." In TM modeling, the whole and the parts are represented as a thimac and subthimacs. The integral nature of the thimac and subthimacs is a duality between thingness and machinery; therefore, aggregation applies to both. Accordingly, the *machines* (of the whole and its parts) are the factors that determine the thimac's structure.

In TM modeling, the *type of flows (activities) of the parts' machines* compared to their whole's machine is the differentiating factor between loosely coupled whole thimacs and a tightly coupled whole to be called *object-oriented* (OO) thimacs. The type of coupling is indicated by the flows (and triggering) of the whole, its parts and the outside of the thimac. A tighter coupling leads to whole/parts integration and the containing thimac's increased control over its parts. The type of control, in this context, refers to UML-like control, in which, say, a subclass overrides a superclass feature by defining a feature with the same name.

In an OO thimac, the whole's machine controls all activities with the outside, in contrast to the general non-OO thimac, in which free flows of interaction with the outside are performed in the whole and its parts. An OO thimac is an internally structured whole whose parts operate as one unit in relation to external flows and other wholes. In such a structure, *objectification* involves superimposing restrictions on a thimac's inside-ness as a "hard wholeness" to convert it to an OO thimac. OO thimacs are characterized by wholeness, which takes control of the parts' *handleablity* (create, process, release, transfer and receive) when interacting with the outside. For simplicity's sake, hereafter, we may refer to *OO thimacs* as simply an *object thimac* and yet a mere *object*. We will refer to the technical object as an *instance* [17]. Instances are distinguished from classes and thimacs, which are interpreted as descriptions of sets of individuals. The distinction between instances and classes is frequently blurred, giving rise to undesirable ambiguities and misinterpretations in modeling areas [18].

In this paper, we explore the notion of aggregation, emphasizing aggregation in UML. In section 2, we briefly review the basics of TM modeling and include two new subsections: *thimacs and objects* and *thimacs and aggregation*. In section 3, we introduce an example that contrasts non-object thimacs and object thimacs with a discussion about the whole-part relationship. In section 4, we discuss the claim that the aggregate has no behavior of its own other than to coordinate its parts' behavior. The rest of the paper involves aggregation samples from the literature of cases of aggregation in UML.

## II. THINGING MACHINE MODEL

According to Evans [1], the conceptual model captures how we choose to think about the domain as we select and break down concepts and relate them. One fundamental concept is the notion of a *thing*. What we call a *thing* "is itself a *process*, a ceaseless coming to be and passing away" [19] (italics added). Such thinking resonates in the TM model, in which things are broadened as *thimacs*. A thimac has a dual nature (Fig. 1): the machine side and the thing side. The machine here acts as a metaphor for a mechanism or apparatus to represent the "mechanical side" of being.

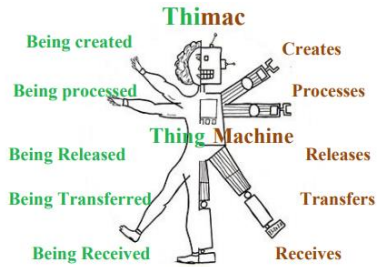

Fig. 1 Thimac duality

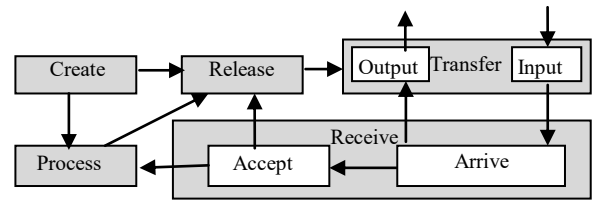

Fig. 2. Thinging machine

*Duality* is a concept that is widely used in philosophy and various branches of special knowledge (in physics, mathematics, chemistry and others) [20]. Distinguishing "between *form and content* and between *process and object*, but, like *waves and particles* in physics, they have to be united in order to appreciate light" [21].

*A. Thinging Machine*

The TM machine has five actions: create, process, release, transfer and receive (Fig. 2). In TM modeling, thingness and machinery cannot be separated, but it is often convenient to focus on one or the other aspect.

A machine's generic TM actions (Fig. 2) can be described as follows:

**Arrive**: A thing moves to a machine.
**Accept**: A thing enters the machine. For simplification, we assume that all arriving things are accepted; therefore, we can combine arrive and accept stages into the **receive** stage.
**Release**: A thing is ready for transfer outside the machine.
**Process**: A thing is changed, handled and examined, but no new thing results.
**Create**: A new thing "comes into being" (is found/manifested) in the machine and is realized from the moment it arises (emergence) in a thimac. Such a description of "being" avoids involving, at least partially, the general philosophical analysis about such a topic. It echoes Quine's slogan, "To be is to be the value of a variable." In such an approach, we should examine the TM model of the world and determine which thimacs *must be* in the domain of the TM description. Such a shaky assumption is applied to the static representation, but this type of existence is tightened at the events level, where every thimac bonds with time.

Note that for simplicity's sake, we omit *create* in some diagrams, assuming the box representing the thimac implies its existence.
**Transfer**: A thing is input into or output from a machine.

Additionally, the TM model includes the triggering mechanism (denoted by a dashed arrow in this article's figures), which initiates a (non-sequential) flow from one machine to another. Multiple machines can interact with each other through the movement of things or through triggering. Triggering is a transformation from movement of one thing to movement of a different thing.

*B. Thimacs and Objects*

Thimacs can be classified into two types, *non-object* and *object*. The assemblages of thimacs are formed from a juxtaposition of subthimacs that are bonded into a structure at a higher level of which they become parts. Thimacs comprise parts, which themselves are thimacs that comprise parts, and so on. Thimacs can't be reduced to their parts, as they have their own machines. The subthimac is a *component* of its thimac and is a proper part of that thimac, which has a boundary that corresponds to discontinuities in reality. All components are parts (subthimacs).

The notion of a thimac does not only include components, e.g., the left side a car (these ideas are based on ideas from [22]). The relationship between a thimac and subthimac can be aggregative (containment), which is a relation that holds when a subthimac (the containee) is located within a thimac, the container [22]. In general, parts of a thimac can interact with the outside of their thimac except when they are object parts (subthimacs) of a thimac. Each subthimac (part) has a freedom of interaction (defined in terms of the five TM actions: create, process, release, transfer and receive) with the containing thimac's interiority and *externality*. The object thimac is a thimac that controls the outside interaction of its parts. Object thimacs can exist because their boundaries define *what is in and out* and determine what can pass (we take this expression from Evans [1] referring to *cells*). The notions of *loose* (parts have independence) *non-object thimacs* vs. fixed *object thimacs* with regard to actions that can provide what is needed in a model that "leaves high-contention points looser and strict invariants tighter" [1].

*C. Thimacs and Generalization*

To relate these ideas to a known structure, consider a thimac and generalization as different types of abstraction [23], where sub-things can be thought of generically as a single named thing. "Generalization" refers to classifying similar things in a generalized manner, e.g., a parent-child relationship, in contrast to the association that links different things. Generalization signifies *kind-of* relationships (inheritances) whereas association expresses *has-a* relationships (commonalities).

According to Smith [23], "Generalization is perhaps the most important mechanism we have for conceptualizing the real world. It is apparently the basis for natural language acquisition." Fig. 3 shows a hierarchy of classes in Smith's generalization (top picture) and the corresponding TM representation (bottom picture). Note that in the TM model, the whole (i.e., thimac) is not identical to the parts that compose it.

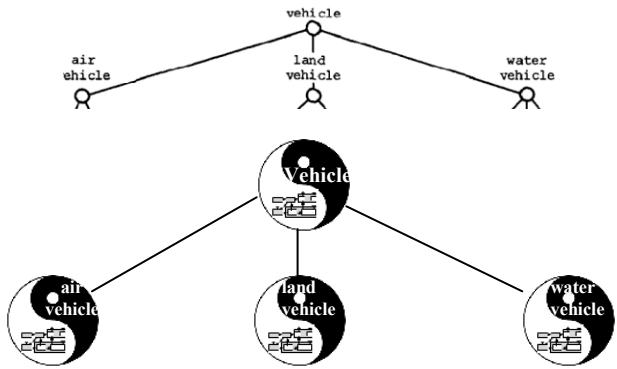

Fig. 3 The so-called generic hierarchy over vehicles (top - partial, from [23]) and the corresponding TM representation (bottom)

(e.g., the whole has its own machine) regardless of whether it has properties. Thus, the TM model brings the notion of aggregation to a higher level, at which *things* and their *machines* (behavior) are aggregated simultaneously. In Smith's example (Fig. 3), according to the TM, not only are air, land and water vehicles aggregated as vehicles; their actions (create, process, release, transfer and receive) are aggregated as holistic actions, e.g., moving a vehicle means moving the whole (air, land and water vehicles), e.g., to a new place, owner, etc. The whole's five TM actions are potentialities of dynamic manifestation of the wholeness.

Changing the *vehicle* from fossil fuel to electrical energy means changing air, land and water vehicles from fossil-fuel-based to electric vehicles. Of course, this does not negate that each part has its own "properties." According to Rajarnoney and Koo [24], *behavioral aggregation* is the process of grouping together a set of individual entities that collectively behave as a unit. In TM, *behavioral* aggregation is the thimac *machine* that represents the behavior of all of its subthimacs as a unit, e.g., release transfer vehicle involves releasing and transferring all instances of water, land and air vehicles. Assuming a strict separation of classes and instances (first-order logic) [18], in a TM, an *instance*, to be discussed later, is a TM *event* (static subdiagram bonded to time).

### III. More on Thimacs and Object Thimacs

An o*bject thimac* is a special type of thimac, in which actions related to communications with the outside are centered in the thimac. In Smith's vehicle hierarchy (Fig. 3 (top)) all nodes are assumed to be objects where second level objects inherent the object's attributes; however, they may have their own attributes. If a subthimac, e.g., air vehicle, is an OO thimac, it has a better-ordered structure regarding its communication (flows and triggering) with other thimacs and its part-whole relations. In UML, if objects are tightly bound by a part-whole relationship, their relation is an aggregation. If they are independent, even though they may often be linked, their relation is an association.

We claim that object thimacs appear in the model description when some of the thimacs manifest their individuality and independence (as identified by the modeler) in such a way that minimizes outside *handling* of their parts (see Fig. 4). In an object thimac, the parts hold together in an assemblage with the containing object thimac as their agent in the interaction with the outside. Therefore, an OO thimac is a thimac with more *organization* that limits flows of its parts' actions.

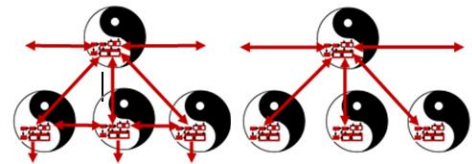

Fig. 4 Non-OO thimac (left) and OO thimac (right)

According to Brinkerhoff [25], the whole-part relationship is often exemplified by the "has a" phrase. But unlike inheritance, which is fundamental to object orientation and implemented with dedicated syntax, the whole-part relationship is implemented with member variables. Inheritance is a special case of an association denoting a "kind-of" hierarchy, in which the children classes inherit the parent class's attributes and operations.

As aforementioned, there are two ways to implement the whole-part relationship, and UML-class diagrams distinguish between them. For example, in both cases, we can read the relationship as *a car has an engine* or *a car has a transmission*. The same relationship can be read in the opposite directions as "part of": *an engine is part of a car* or *a transmission is part of a car* (See Fig. 5) [25]. In composition (referred to in the introduction as "strong aggregation"), the parts are bound strongly or tightly to the whole object. The parts are created and destroyed at the same time that the whole is created and destroyed. In weak aggregation, the part objects are bound weakly or loosely to the whole object - the parts exist outside the whole. The parts and the whole may be created and destroyed at different times.

To understand these relationships, we illustrate the notion of a non-OO thimac (Fig. 6) vs. an OO thimac (Fig. 7) in a TM.

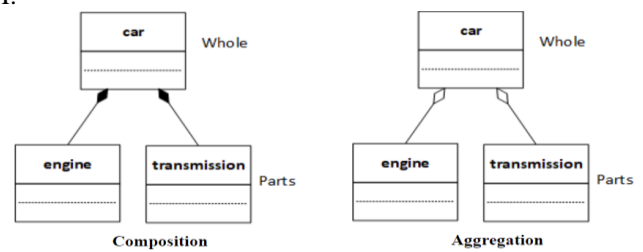

Fig. 5 Composition and aggregation (From [25])

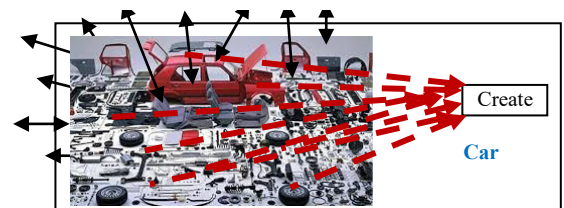

Fig. 6 Non-OO thimac *car* and its parts (image from *https://gomechanic.in/blog/car-spare-parts/*)

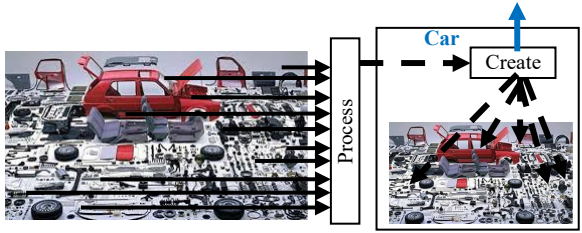

Fig. 7 OO thimac *car* created from parts that have become *its* parts. To manipulate any part, now one must disassemble it from the object car.

In Fig. 7, the presence of all parts of a car triggers the creation of a car thimac even though the actual car has not been assembled. Fig. 8 further illustrates this phenomenon with a table that emphasizes looseness. We recognized that the four pieces in Fig. 8 can form a table, thus constructing the table. Aristotle introduces such a primitive distinction between the two aggregations because they differ in the way they are unified and in the way they can be composed and decomposed [26]. However, in contrast to Aristotle, in a TM, entities (thimacs) are not mere sums of their components because the whole always has its machine. In Fig. 8, the parts are still loose and can be manipulated as independent components.

Note that for simplicity's sake, we do not show the release, transfer and receive actions. This conceptualization step is a *thinging stage*, at which the domain comprises thimacs that emerge or are thrown toward us (things that emerge from the world). The thimacs have a minimum level of organization (conceptual space) with respect to the whole-part relationship and their behavior in terms of the five generic actions.

Fig. 9 shows that subthimacs create the *object* thimac car, in which the subthimacs are now parts of the object. In terms of the TM-event model, first, subthimacs are processed to create an object with parts.

Returning to Brinkerhoff's example [25], Fig. 10 shows the composition of engine and transmission as an OO car. The engine and a transmission move (transfer, receive) as integral parts of the car. If it is necessary to remove any part, the whole car is processed as Fig. 11 shows. Note that the processing triggers *transfer/receive* because engine and transmission were *already* created but "hidden" inside the car. Transfer/receive here entails an *appearance*.

## IV. AGGREGATION EXAMPLE

According to the CORBA (Common Object Request Broker Architecture) [7], aggregation describes the assembly of objects to create a new object. When the aggregate (interface) is invoked, it sends instructions to its parts to implement the requested function.

The aggregate has no behavior of its own other than to coordinate its parts' behavior. For example, a car comprises an engine, transmission, fuel system, and other parts. When drivers want to move the car forward, they use the car's interface, that is, the ignition switch, the gas pedal, the brake pedal, and so on, rather than accessing the parts directly.

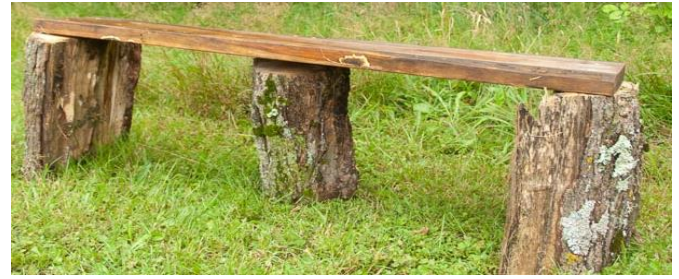

Fig. 8 Illustration of looseness in composition (image from Google)

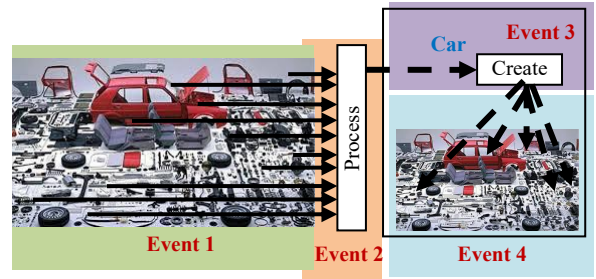

Fig. 9 Events that lead to creating an OO thimac, *car*

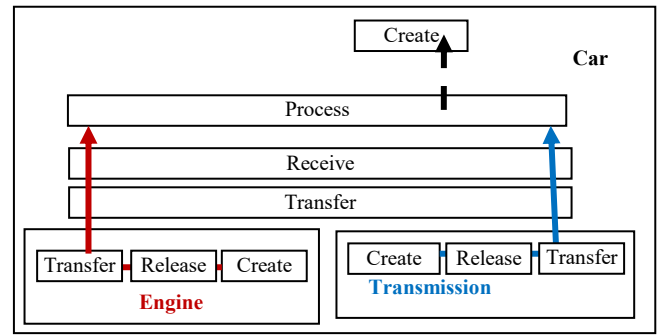

Fig. 10 TM representation of the OO thima *car*

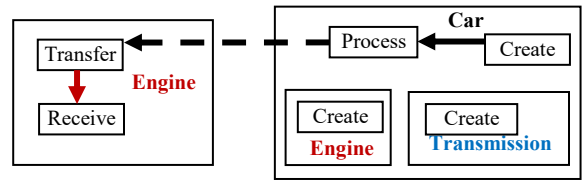

Fig. 11 Removing engine

For example, when the gas pedal is pressed, the fuel system pumps gas, the engine turns over and the transmission responds to the speed changes. The intent of aggregation is to hide a complex object's makeup by requiring client objects to talk to the aggregate [7].

### A. TM Static Model

Fig. 12 shows the static TM model that corresponds to the CORBA [7]'s car example. First, a signal flows to the engine (blue number 1) to turn it on (2). The same process occurs in the transmission and fuel system to create the car's movement (7). This movement is processed in the sense that it continues (8).

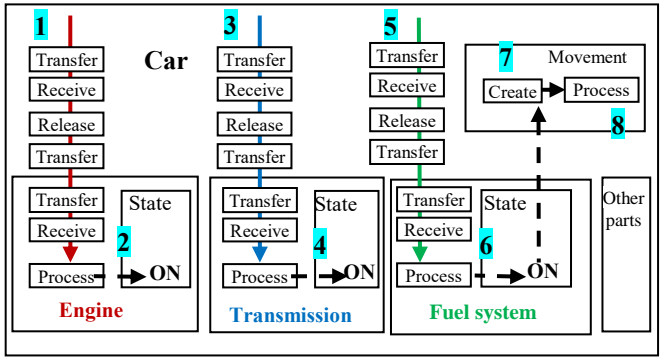

Fig. 12 Static TM model of the car example

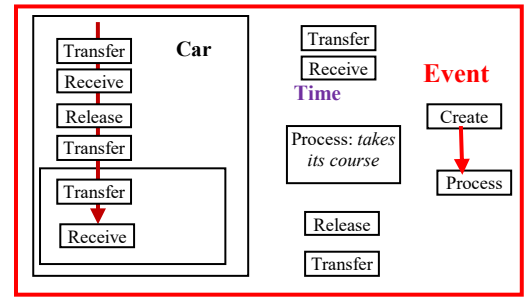

Fig. 13 The event: *A signal flows to the car (e.g., from the ignition switch —included in the model) which sends it to the engine*

### B. Event Model

An event in a TM involves a subdiagram of the static model bonded with a time subthimac to form an event thimac. Fig. 13 shows the event *a signal flows to the car (e.g., from the ignition switch, not included in the model), which sends it to the engine*. For simplicity's sake, we will use only the subdiagrams of the static model to represent events. For example, Fig. 14 shows selected events based on Fig. 12. Initially, each of the generic actions can be an event; however, for human understanding, these events form a set of larger events. Accordingly, Fig. 14 presents the selected events, which can be described as follows:

E1: A signal flows to the car (e.g., from the ignition switch, included in the model), which sends it to the engine.
E2: The signal is processed, causing the engine to turn on.
E3: A signal flows to the car, which sends it to the transmission.
E4: The signal is processed, causing the transmission to change position.
E5: A signal flows to the car, which sends it to the fuel system.
E6: The signal is processed, causing the fuel system to supply fuel.
E7: The car initiates movement.
E8: The car continues moving.

Fig. 15 shows the behavioral model.

### C. Data Model

The CORBA's [7] example of a car is meant to develop a data model that helps describe the structure of the data elements in an information system and the relationships between data elements. The structure forms the basis of the physical data model. To give an example of how to use it, the TM representation in data modeling, let us assume that *car* is an OO thimac and that *engine*, *transmission* and *fuel system* are attributes with primitive data types, e.g., engine serial number = DJ51279, transmission type = automatic, fuel system type = sequential injection. Fig. 16 shows the car's *construction* (e.g., construct in C++) from the attributes value (a tuple in relational database). To save space, because the static model is embedded in the events model, only the event model is presented. Because the car is an OO thimac, filling the attributes values goes through *car*. Fig. 17 shows the behavioral model.

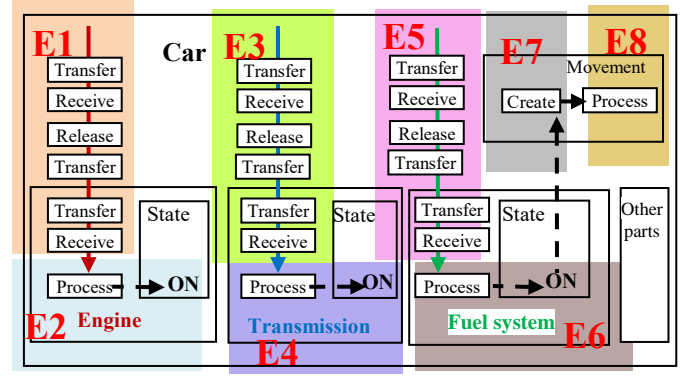

Fig. 14 Events model of the car example

$$E1 \rightarrow E2 \rightarrow E3 \rightarrow E4 \rightarrow E5 \rightarrow E6 \rightarrow E7 \rightarrow E8$$

Fig. 15 Behavioural model of the car example

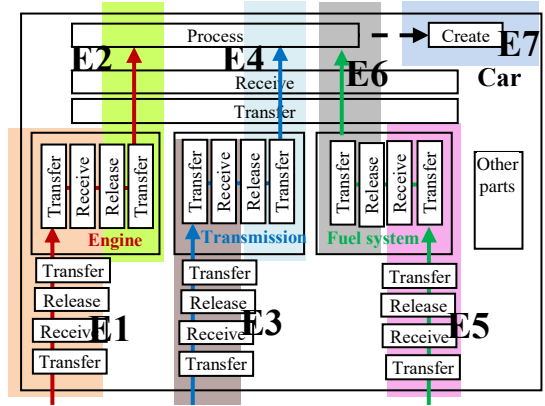

Fig. 16 Events model of the construction of a car object.

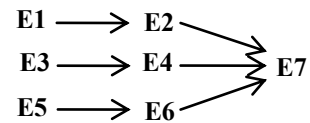

Fig. 17 Behavioural model that corresponds to Fig. 16

## V. POLYGONS AND CIRCLES

Fig. 18 shows an example of aggregation and composition from Fowler [8]. The compositions to *point* indicate that any instance of point may be in either a *polygon* or a *circle* but not both. Many polygons and circles, however, may share an instance of *style*. Furthermore, this implies that deleting a polygon would cause its associated points to be deleted but not the associated style [8].

Fig. 19 shows the TM static model of this *point* (purple number 1) / *style* (2) example. First, we assume that *point* is an OO thimac and style is a non-OO thimac. Polygon and circle are sub-thimacs of point. Therefore, we have two operations that will be illustrated only with respect to *polygon* because the *circle* has similar operations.

### A. Creating a Polygon

As the figure shows, a *point* is created (e.g., by processing a user's request), and it flows to *polygon* (4) according to the user's selection of a polygon. We assume that the polygon's *side* is also given; therefore, the point and the side are processed (6) to create a polygon (7). This instance of the polygon is added to the set of *polygons* (8). *Polygons* (with an *s*) is the set of polygons created in the system. We have not elaborated on how to add a new polygon to polygons, simplifying such a task as processing the current set of polygons and the new polygon (9) to produce a new set.

A similar process is performed to create a circle.

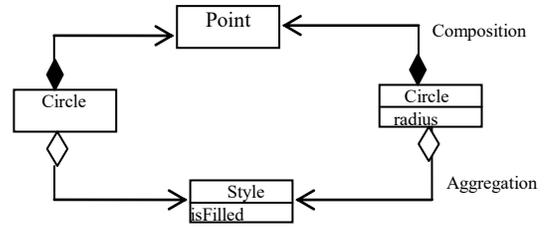

Fig. 18 Aggregation and composition (redrawn, incomplete, from [8])

Note that the polygon and circle are inside a point (the thick blue rectangle). On the other hand, the style as a general thimac includes the polygon and the circle.

### B. Filling/Not Filling a Polygon

*To fill* or *not fill* the polygon, *style* receives an input (10) specifying *fill* or *not fill*, which moves to *point* (11). This step is necessary because the point is an object and an object controls all parts of the project.

- Additionally, *style* is received as an input, the polygon's *point* (12). The fill/not fill value also moves to the point (13). Point sends the style and point values to polygon (14 and 15). There, the point and the polygons are processed (16) to produce an instance of a polygon that is indicated by the given point. (17).

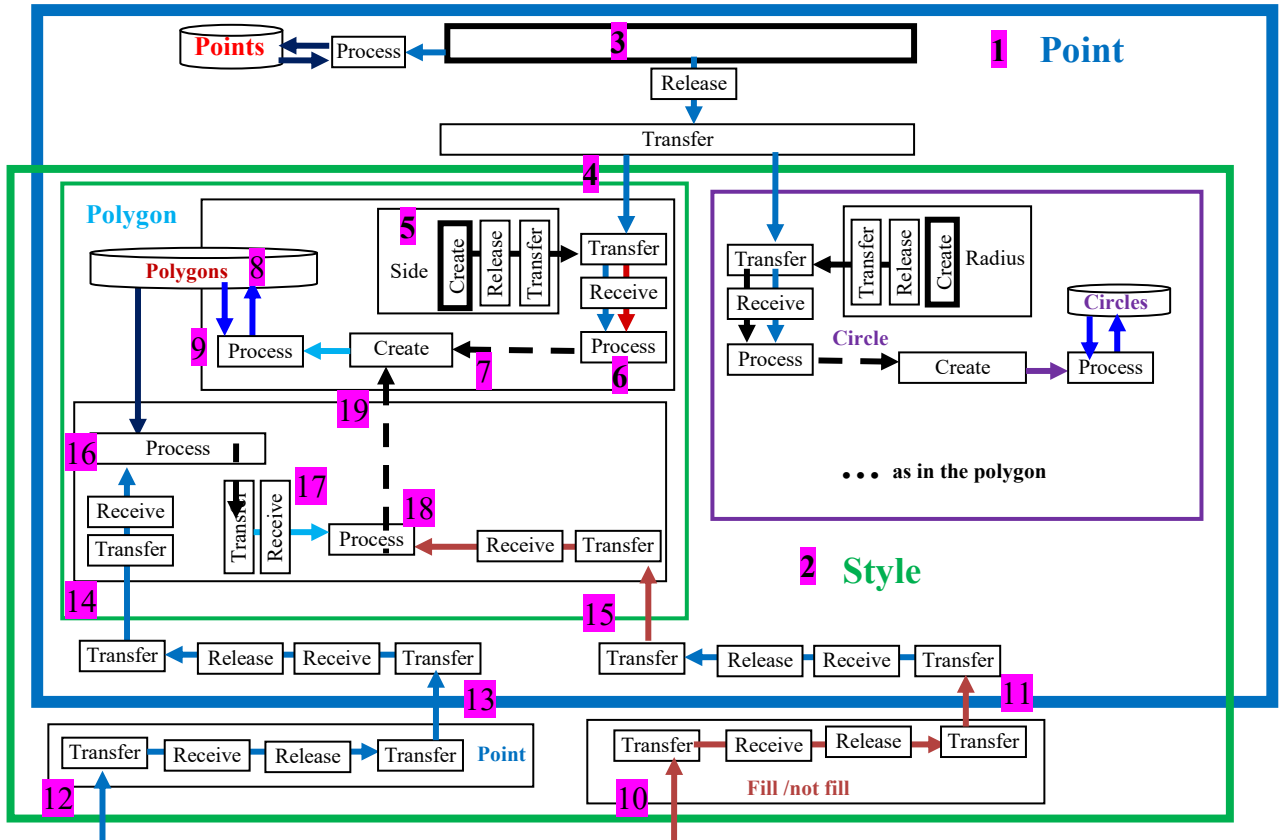

Fig. 19 The static TM model of the *point/style* example

- This instance, along with the style value, is processed (18) to produce a new instance of the polygon (19) that is added to *polygons*, replacing the old version. This adding and replacing process is not elaborated in the figure. It is obvious that the point represents a UML class, and attributes include point value, side and radius with methods including
*Create Polygon (point, side)*
*Create Circle (point, radius)*
*Fill (Polygon, point, fill-notFill value)*
*Fill (circle (point, fill-notFill value)*

At this conceptual level, the objects' construction is left out. Construction of a given object involves building it up from the parts that it contains. This is a composition operation that takes the parts into the whole object using the TM operations (create, process, release, transfer and receive).

At the implementation level, a polygon or circles can be constructed using pointers to the inherited point value. The TM representation provides a complete description and control for projects and not just iconic representation, as in the case of UML classes. The control inside an object is clearly specified in terms of flows

Fig. 20 shows the event model for the two operations above, and Figs. 21 and 22 show the behavioral models.

*C. Deletion*

We still must specify the deletion operation of polygon or circle that would cause its associated *point* to be deleted, as Fowler [8] mentioned. Deleting a *polygon* would cause its associated *point* to be deleted but not the associated *style*. Deleting a polygon requires specifying its point and side. Several polygons could be centered on a given point. Accordingly, the delete request is received in point, as Fig. 23 shows (pink number 1). The deletion request is processed (2) to extract the point and side values (3 and 4, respectively).

The point value (5) and *points* (6) are processed (7) to extract the data about the object point. The data in points is a set of objects that includes, for each point, object point value in additional to data to "link" the polygon to a polygon object.

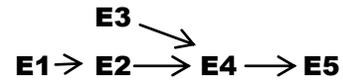

Fig. 21 The behavioural model of creating a polygon

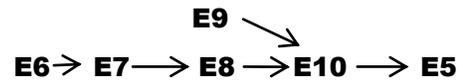

Fig. 22 The behavioural model of adding fill/not-fill to a polygon

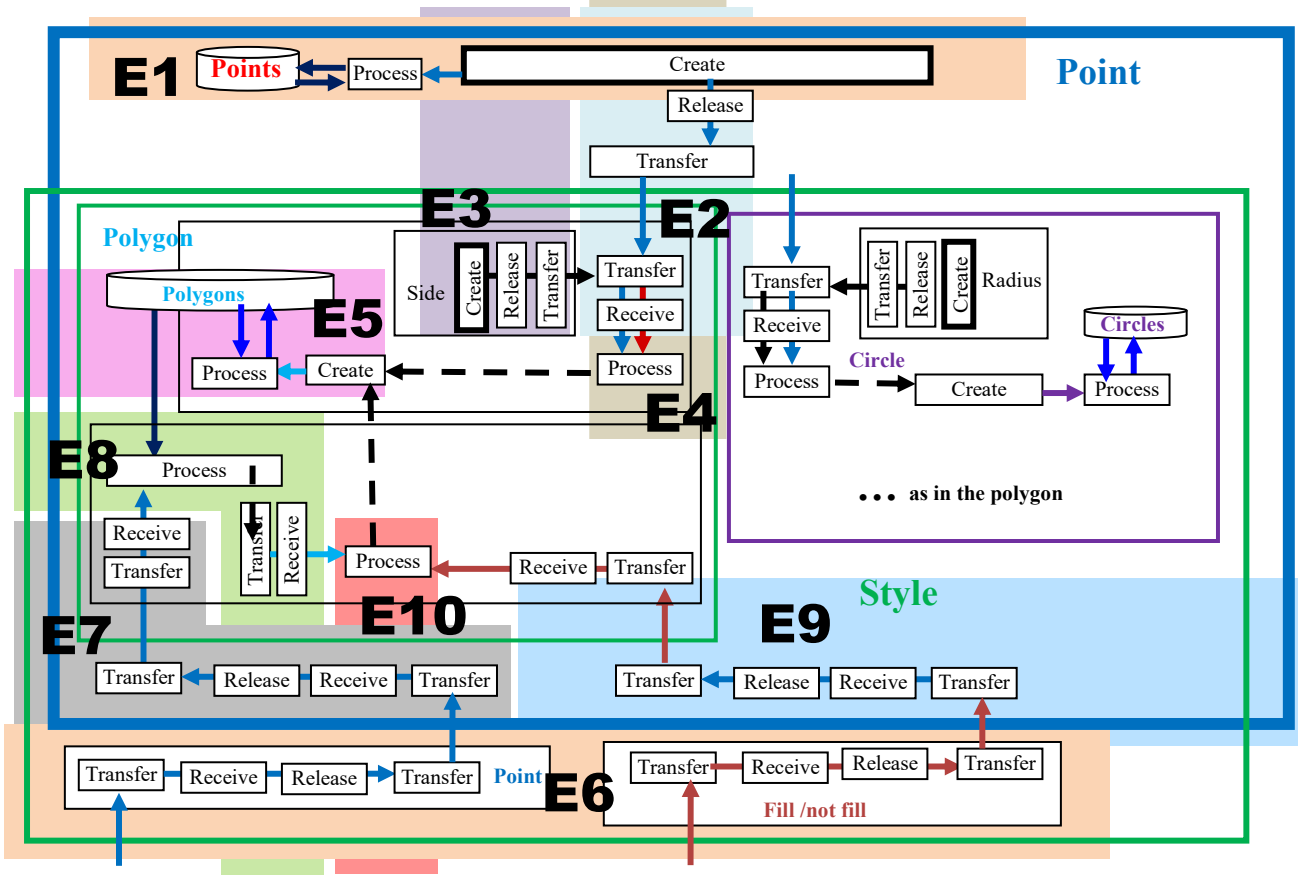

Fig. 20 The events model of the *point/style* example

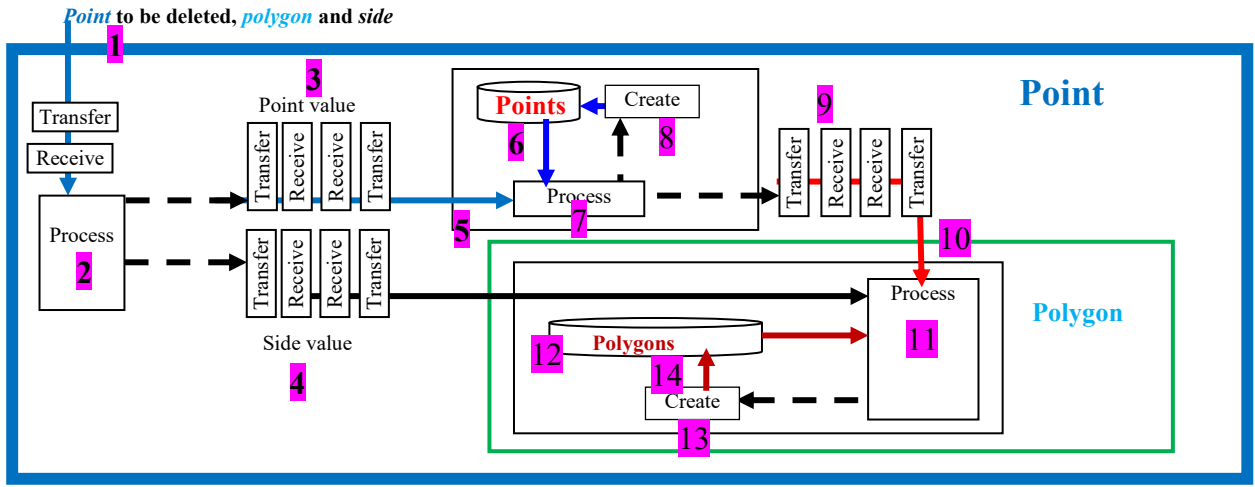

Fig. 23 Static description of deleting a point given its *type* and *side*

This process results in
- Creating points (8) that do not include the record of the deleted polygon.
- Extracting and sending the point record of the deleted polygon instance to polygon (9 and 10).

In polygon, the *points* record of the deleted polygon is processed (11) along with the *polygon*s set (12) to create (13) a new set of polygons (14). Accordingly, the deletion operation of polygon or circle objects would cause the deletion of their associated *point* object.

## VI. POLYGONS AND CIRCLES

Wagner and Diaconescu [27] studied aggregation as a special form of a part-whole association, where the parts of a whole can be shared with other wholes. This is illustrated by the aggregation between the classes *DegreeProgram* and *Course*, as Fig. 24 shows.

A course is part of a degree program and can be shared among two or more degree programs. The assumption is that DegreeProgram and Course are objects. That means, in TM terms, that they are handled as a whole. To illustrate the TM modeling, we will focus on the operation of *adding a course* to a DegreeProgram as a request from the user.

### A. Static Model

Fig. 25 shows the TM static model of this example.

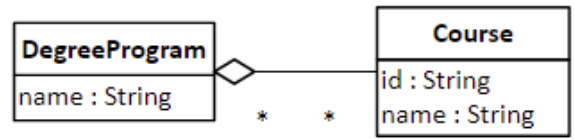

Fig. 24 Example from [27]

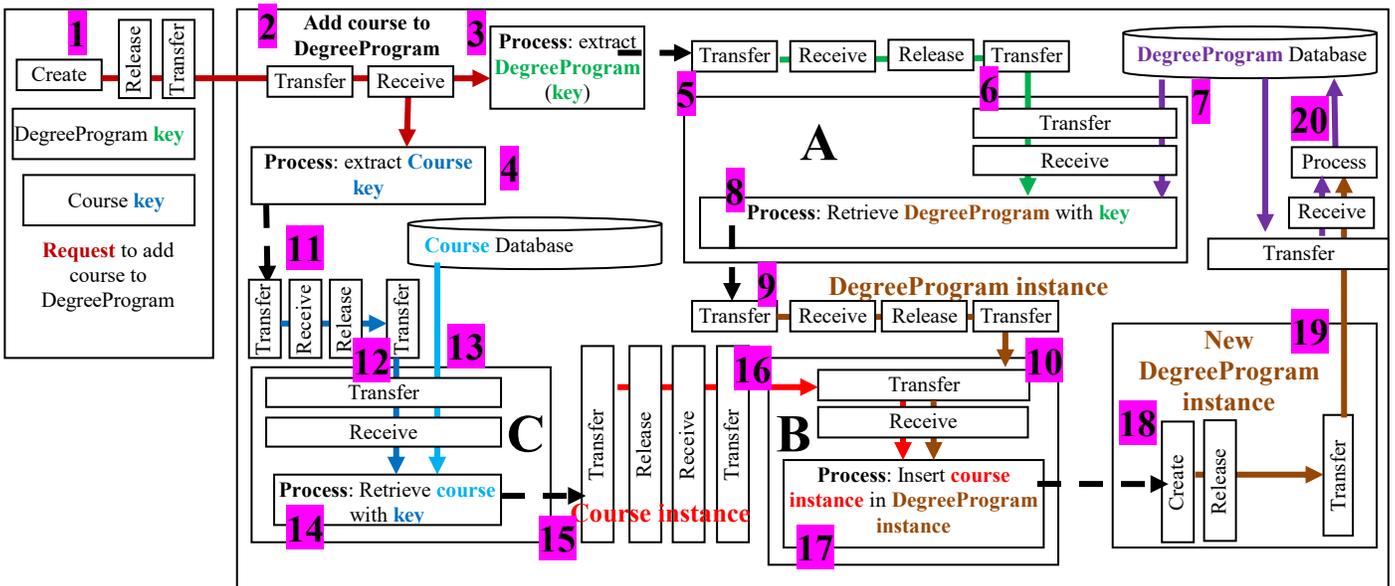

Fig. 25 The static TM model of the *Course/DegreeProgram* example

We assume that the request has two identifiers: a key of Course and key of DegreeProgram. Additionally, we assume that the request is an object; therefore, the two keys are input and handled together instead of each one being input and handled separately. We also assume that a request to add a course to DegreeProgram is created that includes DegreeProgram and the course key (see pink number 1 in the figure). The request is received at the module in charge of the process of adding a course to a DegreeProgram. (2). There, the two keys are extracted from the request: (a) the DegreeProgram key (3) and (b) the course key (4).

The process of extracting the DegreeProgram key triggers the arrival (transfers and receive) of the key (5), which flows to module A (6). Additionally, the DegreeProgram database is made available to A (7). Processing in A would retrieve the instance in DegreeProgram with the given key (8). This operation results in the availability (9) of that instance of DegreeProgram, and the instance flows to another module, called B (10).

The same thing happens to the course key (11), which flows to module C, where the *course database* is available (13). When the Course key and database are processed (14), a course instance is extracted (15), which flows to B. In B, the course and DegreeProgram are processed (17) to create a new DegreeProgram instance (18), which is processed (19) in the DegreeProgram database, replacing the old instance (20).

### B. Association Between Course and DegreeProgram

Fig. 26 shows the event model of this example, and Fig. 27 shows the behavioral model. If we are interested in the association between Course and DegreeProgram, we can eliminate some technicalities to produce Fig. 28.

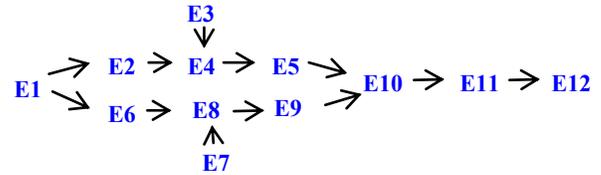

Fig. 27 The behavioural TM model

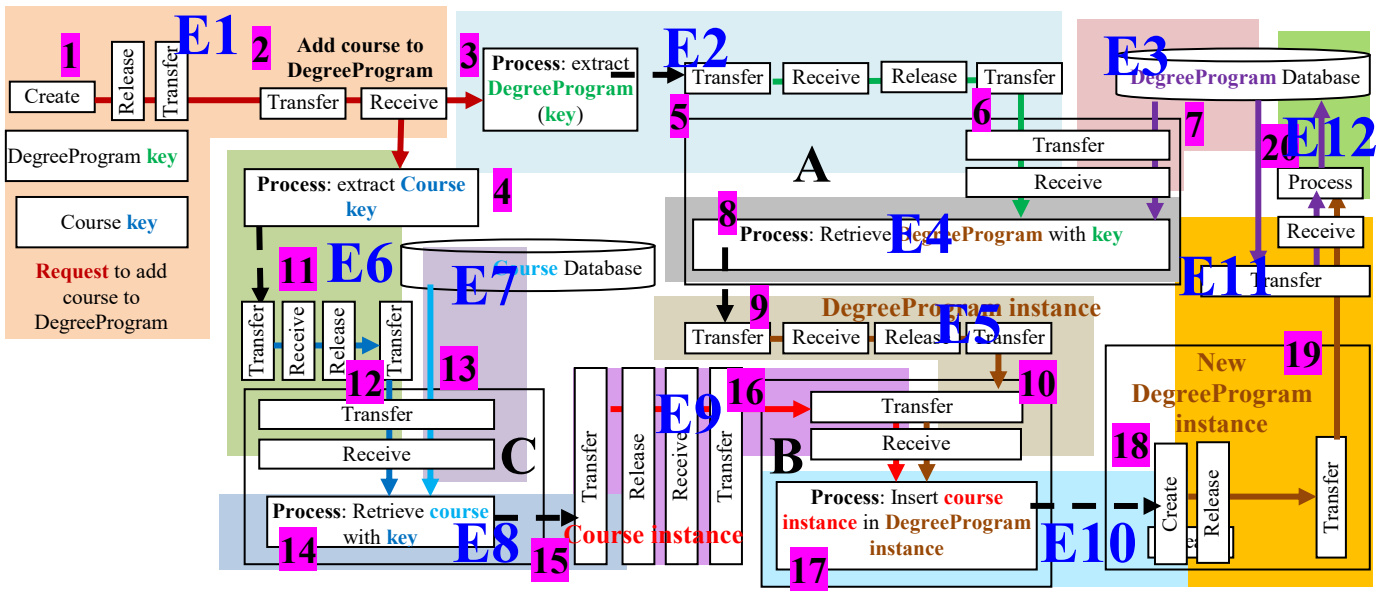

Fig. 26 The events model of the *Course/DegreeProgram* example

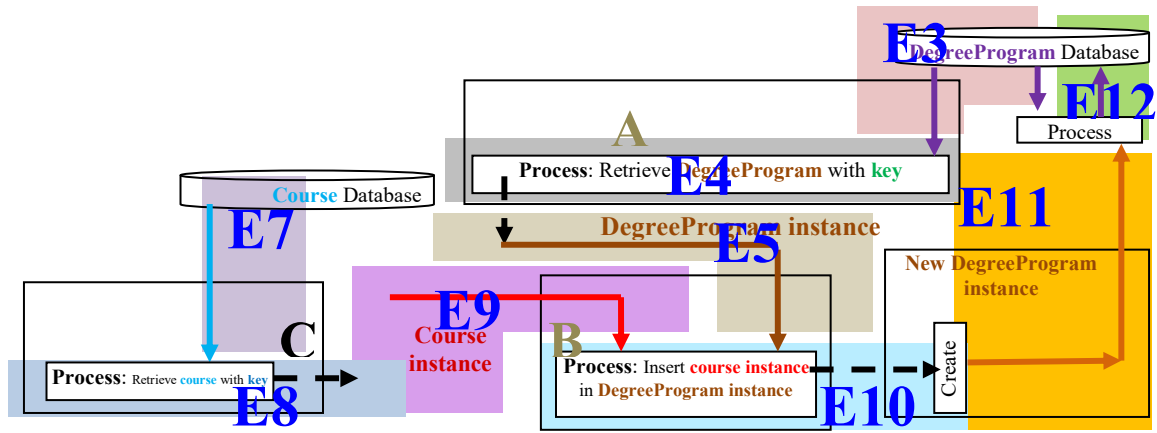

Fig. 28 Events connecting Course and DegreeProgram

Accordingly, we note that the basic association between the two objects, Course and DegreeProgram, is accomplished by linking (e.g., keys) instances of Course inside DegreeProgram. Consequently, we can represent the flow events from Course to DegreeProgram as in Fig. 29. The events in Fig. 29 are merged to produce Fig. 30. Fig. 31 shows the reoccurrence of the combined event that reflects the minimum representation of the association comparable to the UML class diagram in Fig. 32.

Note that the many-many relationship is implicit in the TM diagram of inserting any Course into any DegreeProgram without *constraint*s. Additionally, the Course is an independent object with respect to DegreeProgram. In this case, when one deletes an instance of Course, they should consider the Course entries in DegreeProgram.

VII. CONCLUSION

As stated in the introduction, this paper's purpose is to explore the possibility of refining the notion of aggregation in the context of conceptual modeling analysis, emphasizing aggregation in UML using the TM model as a tool in such a venture. We have done so by contributing to ontological conceptual clarity about aggregation through the notion of a thimac (thing/machine) as a building block for the specification domain.

One achievement is extending the notion of aggregation to behavioral aggregation, in which individual entities collectively behave as a united thimac, which by definition has its own machine. This machine-based approach seems promising compared to the ontological positions that are based on substance or relation. Now, the concept of a thing/machine seems to substitute for traditional terms of "substance," "body," "object" and "thing." The notion of an object thimac presents a hierarchy of organizational sophistication of control through actions.

Another accomplishment is showing, initially, that the TM model provides a tool to represent the details of relationships that are embedded in the aggregation as a type of association in which objects are configured together to form a more complex object.

Still, a great deal of research is needed to provide a better understanding of the notion of aggregation and the whole-part relationship that will be pursued in future works.

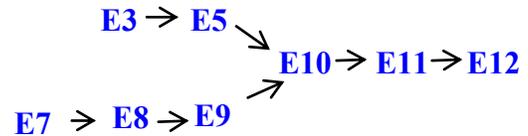

Fig. 29 Basic events in Fig. 28

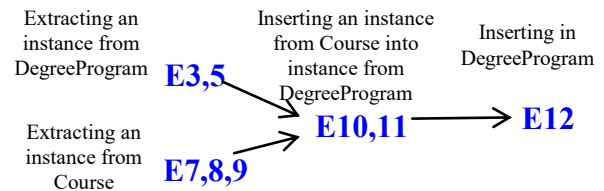

Fig. 30 Merging basic events in Fig. 29

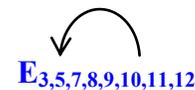

Fig. 31 Reoccurrence of the combined event

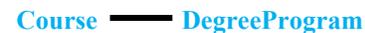

Fig. 32 Bipartite relation